\documentclass[10pt,aps,twocolumn,secnumarabic,balancelastpage,amsmath,amssymb,nofootinbib,hyperref=pdftex,prl,citeautoscript]{revtex4-1}

\usepackage{graphics}
\usepackage[pdftex]{graphicx}
\usepackage{tabularx}
\graphicspath{{/Users/chrisreeg/Desktop/Doctoral_Research/My_Papers/"Proximity Gap"/Figures/}}
\usepackage{longtable}
\usepackage{amsmath}
\usepackage{color,soul}

\makeatletter
\def\p@subsection{\thesection\,}
\makeatother

\newcommand{\bo}[1]{{\bf #1}}

\newcommand{\kp}{k_\parallel}

\usepackage{hyperref}
\hypersetup{colorlinks=true,linkcolor=blue,anchorcolor=blue,citecolor=blue,filecolor=blue,urlcolor=blue,bookmarksnumbered=true,pdfview=FitB}

\begin{document}

\title{Hard gap in a normal layer coupled to a superconductor}
\author{Christopher R. Reeg}
\author{Dmitrii L. Maslov}
\affiliation{Department of Physics, University of Florida, P. O. Box 118440, Gainesville, FL 32611-8440, USA}
\date{\today}
\begin{abstract}
The ability to induce a sizable gap in the excitation spectrum of a normal layer placed in contact with a conventional superconductor has become increasingly important in recent years in the context of engineering a topological superconductor. The quasiclassical theory of the proximity effect shows that Andreev reflection at the superconductor/normal interface induces a nonzero pairing amplitude in the metal but does not endow it with a gap. Conversely, when the normal layer is atomically thin, the tunneling of Cooper pairs induces an excitation gap that can be as large as the bulk gap of the superconductor. We study how these two seemingly different views of the proximity effect evolve into one another as the thickness of the normal layer is changed. We show that a fully quantum-mechanical treatment of the problem predicts that the induced gap is always finite but falls off with the thickness of the normal layer, $d$. If $d$ is less than a certain crossover scale, which is much larger than the Fermi wavelength, the induced gap is comparable to the bulk gap. As a result, a sizable excitation gap can be induced in normal layers that are much thicker than the Fermi wavelength.
\end{abstract}

\maketitle

\begin{figure}[t!]
\includegraphics[width=\linewidth]{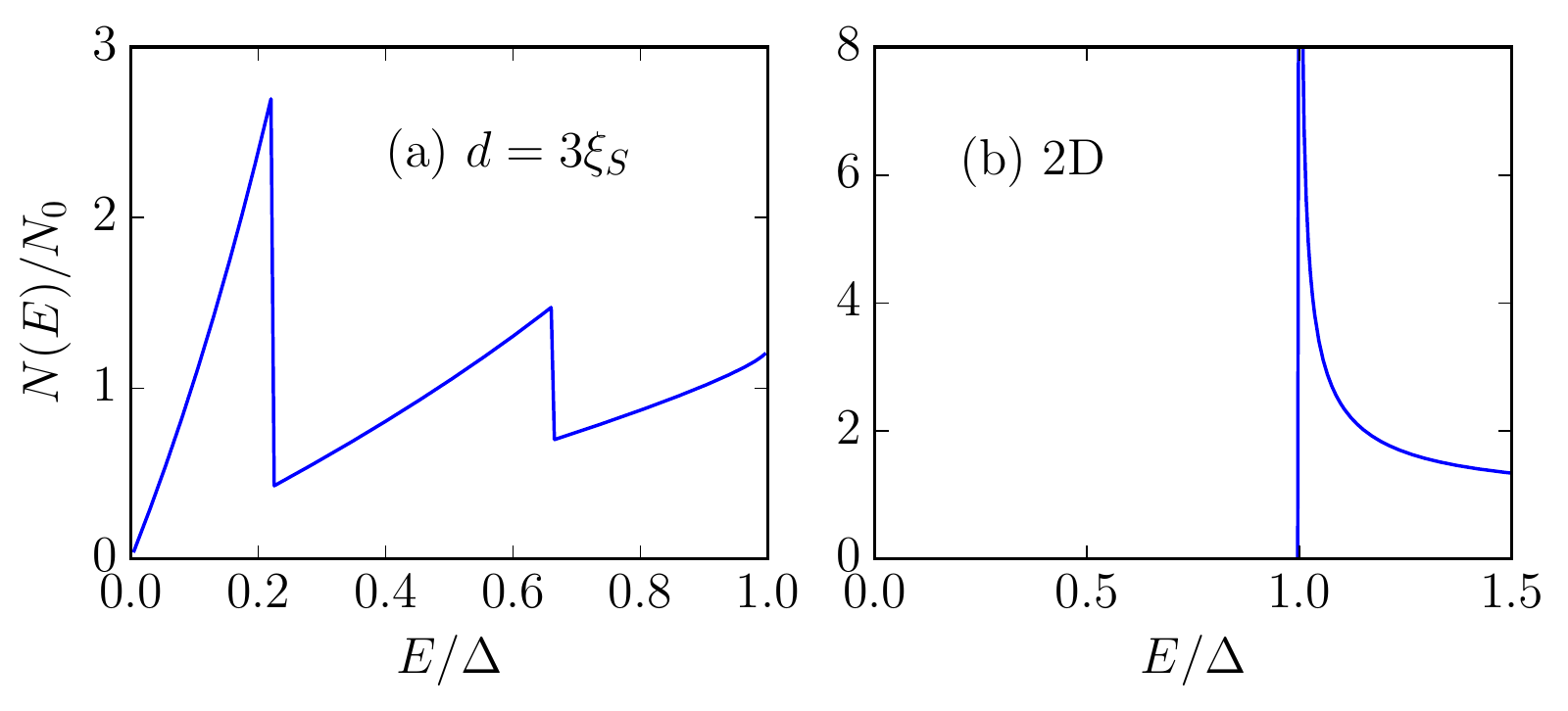}
\caption{\label{DOScomp} (Color online) (a) A quasiclassical result for the density of states $N(E)$ in a normal layer coupled to a superconductor. In this approximation, one obtains a gapless density of states that vanishes linearly at the Fermi energy. (b) A tunneling-Hamiltonian result for the density of states in a two-dimensional (2D) normal layer coupled to a superconductor. A BCS-like gap is induced, with the size of the gap determined by the transparency of the SN interface (shown here for a highly transparent interface).}
\end{figure}

\paragraph*{Introduction.}
There are two seemingly distinct paradigms for understanding the superconducting proximity effect. In a more traditional approach based on the quasiclassical theory \cite{Andreev:1964,Eilenberger:1968} (which we dub ``mesoscopic"), Andreev reflection gives rise to a nonzero pairing amplitude but does not induce a superconducting gap in a clean normal layer \cite{deGennes:1963} [see Fig.~\ref{DOScomp}(a)]. This seems to stand in stark contrast to the approach adopted in more recent studies of the proximity effect in materials that are a single atom thick. In this approach (which we dub ``nanoscale''), the tunneling of Cooper pairs opens a gap in the excitation spectrum of the layer, and this gap can be as large as the bulk gap of the superconductor ($\Delta$) \cite{Sau:2010prox,Potter:2011,Kopnin:2011,Takane:2014,Alicea:2012} [see Fig.~\ref{DOScomp}(b)]. 
 The latter approach has become increasingly important in recent years, owing to the intense push to realize Majorana fermions in condensed matter systems \cite{Alicea:2012,Leijnse:2012,Beenakker:2013}. As topological superconductivity requires the presence of a sizable proximity-induced gap to protect the zero-energy Majorana modes, this aspect of the proximity effect is crucial to the success of any proposal to engineer the topological phase \cite{Fu:2008,Lutchyn:2010,Oreg:2010,Sau:2010,Cook:2011,NadjPerge:2013,Chang:2015,Kjaergaard:2016,Zhang:2016}.

\begin{figure}[t!]
\includegraphics[width=0.7\linewidth]{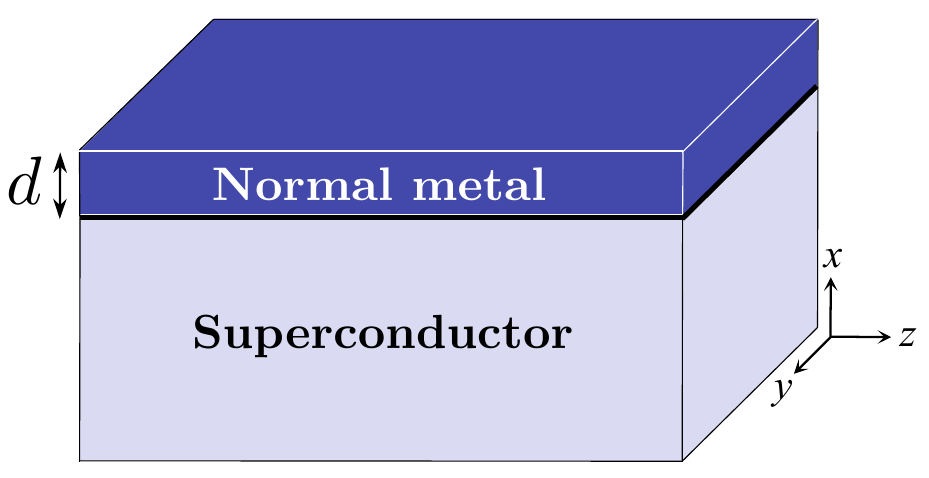}
\caption{\label{geometry} (Color online) (a) Geometry under consideration in this paper, where a normal layer of finite thickness $d$ is placed in contact with a semi-infinite superconductor (both materials are infinite in the $yz$-plane). A sharp potential barrier is included at the SN interface, which is located at $x=0$. }
\end{figure}

In this paper, we attempt to bridge the gap between these two views of proximity-induced superconductivity by studying the evolution of the induced superconducting gap as the thickness of the normal layer ($d$) is changed (see Fig.~\ref{geometry}). In order to treat both mesoscopic and nanoscale systems, we formulate our approach in a fully quantum-mechanical way. We first show that the gapless state of the mesoscopic approach is an artifact of the quasiclassical approximation. Within the same model as in Ref.~\cite{deGennes:1963}, we show that there are two competing energy scales, $\Delta$ and $1/md^2$, that determine the size of the proximity-induced gap ($m$ is the effective mass in the normal layer, and we set $\hbar=1$). The quasiclassical approach misses the latter scale, and we show that a finite gap is induced for any finite $d$. By allowing for arbitrary thickness, we are able to show that for a sufficiently thin junction with $d\lesssim d_c$, the induced gap constitutes a sizable fraction of the bulk superconducting gap. For an ideal junction (no Fermi surface mismatch and no interfacial barrier),
\begin{equation}\label{dc}
d_c=\sqrt{\xi_S\lambda_F},
\end{equation}
where $\xi_S$ is the superconducting coherence length and $\lambda_F$ is the Fermi wavelength. If the layer is metallic, then we always have $\xi_S\gg \lambda_F$, and it is possible to induce a sizable gap in a normal layer that is many atomic layers thick. If the layer is semiconducting but still $\xi_S\gg\lambda_F$, a sizable gap can be induced in a layer that is not in the 2D limit. For example, a sizable gap can be induced in multilayer graphene (i.e., one does not need a monolayer to induce the gap) or in topological insulator thin films. Finally, we address the effects of Fermi surface mismatch and an interfacial barrier, both of which weaken the proximity effect.

Before continuing with our analysis, we must address some overlap between this work and the existing literature. First, we note that Ref.~\cite{BarSagi:1977} obtained a gapped state {\em within} the quasiclassical theory. However, this result is in contradiction with that of Ref.~\cite{deGennes:1963}, which predicts only a gapless state, and we show below that the induced gap is indeed missed by the quasiclassical approximation. Second, we note that Refs.~\cite{Volkov:1995,Fagas:2005,Tkachov:2005} studied the proximity effect in a quasi-2D quantum well, where only the lowest transverse subband is occupied and where the quantum well and superconductor are only weakly coupled. Our model allows us to treat both arbitrary thickness and arbitrary coupling between normal layer and superconductor, and our results coincide with those of Refs.~\cite{Volkov:1995,Fagas:2005,Tkachov:2005} in the appropriate limits.

\paragraph*{Model.} \label{Model}

We consider an SN junction as shown in Fig.~\ref{geometry}, where the normal layer has a finite thickness $d$. We allow the mass $m(x)$, the Fermi energy $E_F(x)$, and the pairing potential $\Delta(x)$ to vary in a stepwise manner across the SN interface. Specifically, we take $m(x)=m_N\theta(x)+m_S\theta(-x)$, $E_F(x)=E_{FN}\theta(x)+E_{FS}\theta(-x)$, and $\Delta(x)=\Delta\theta(-x)$. We also allow for an interfacial barrier of the form $U(x)=U\delta(x)$. Our model is described by the standard BdG equation:
\begin{equation} \label{BdG}
\left[\mathcal{H}_0\hat\tau_3+\Delta(x)\hat\tau_1\right]\psi(\kp,x)=E\psi(\kp,x),
\end{equation}
where $\bo{k}_\parallel$ is the (conserved) momentum in the plane of the SN interface, $\mathcal{H}_0=-\partial_x\bigl[\partial_x/2m(x)\bigr]+k_\parallel^2/2m(x)-E_F(x)+U(x)$, and $\hat\tau_i$ are the Pauli matrices. Because we are interested in studying the induced gap in the normal layer, which should not exceed the bulk gap of the superconductor, we consider only energies $E<\Delta$.

On the superconducting side, we must ensure that the solution to Eq.~(\ref{BdG}) decays into the bulk. On the normal side, we account for the outer boundary by requiring the wave function to vanish at $x=d$. The wave function in the two regions can then be expressed as
\begin{subequations} \label{sol}
\begin{gather}
\psi_S=c_1e^{-ip_+x}\left(\begin{array}{c} u_0 \\ v_0 \end{array}\right)+c_2e^{ip_-x}\left(\begin{array}{c} v_0 \\ u_0 \end{array}\right),  \label{solS} \\
\psi_N=c_3\sin\bigl[k_+(d-x)\bigr]\left(\begin{array}{c} 1 \\ 0 \end{array}\right)+c_4\sin\bigl[k_-(d-x)\bigr]\left(\begin{array}{c} 0 \\ 1 \end{array}\right),
\end{gather}
\end{subequations}
where $u_0^2=(1+i\Omega/E)/2$ and $v_0^2=(1-i\Omega/E)/2$ are the usual BCS coherence factors and $\Omega^2=\Delta^2-E^2$. The momenta defined in Eq.~(\ref{sol}) are given by
\begin{subequations} \label{momenta}
\begin{align}
p_\pm&=k_{FS}\sqrt{\varphi_S^2\pm i\Omega/E_{FS}}, \\
k_\pm&=k_{FN}\sqrt{\varphi_N^2\pm E/E_{FN}},
\end{align}
\end{subequations}
where $k_{F}=2mE_{F}$ is the Fermi momentum and $\varphi^2=1-k_\parallel^2/k_{F}^2$ parameterizes the quasiparticle trajectory.

The boundary conditions to be imposed at the SN interface can be obtained by direct integration of Eq.~(\ref{BdG}) over a narrow region near $x=0$; they are
\begin{subequations} \label{BCs}
\begin{gather}
\psi_N(\kp,0)=\psi_S(\kp,0), \\
\frac{1}{m_N}{\partial_x}\psi_N(\kp,0)-\frac{1}{m_S}{\partial_x}\psi_S(\kp,0)=2U\psi(\kp,0).
\end{gather}
\end{subequations}
The boundary conditions form a set of four coupled equations that must be solved simultaneously. The condition for the solvability of this system of equations determines the excitation spectrum of the SN junction; i.e., a given energy $E$ belongs to the spectrum only if there exists a choice of $k_\parallel$ for which the solvability condition is satisfied. By determining which energies are absent from the spectrum, we can determine the size of the gap that is induced in the normal layer.

\paragraph*{Breakdown of the quasiclassical approximation.} \label{QCsec}
As first shown in Ref.~\cite{deGennes:1963} [and as displayed in Fig.~\ref{DOScomp}(a)], the quasiclassical theory gives a normal layer density of states that vanishes linearly at the Fermi energy. To reproduce the quasiclassical results of Ref.~\cite{deGennes:1963}, we neglect the effects of a sharp interface by setting $U=0$ and by assuming that there is no Fermi surface mismatch between the superconductor and normal layer. The quasiclassical approximation corresponds to expanding the momenta of Eq.~(\ref{momenta}) in the limit
\begin{equation} \label{QCcond}
\varphi^2\gg\Delta/E_F,
\end{equation}
which means grazing trajectories with $k_{||}\approx k_F$ are excluded. This gives
\begin{subequations} \label{QCexp}
\begin{align}
p_\pm&=k_{F}|\varphi|\pm i\Omega/v_F|\varphi|, \label{expansion1} \\
k_\pm&=k_F|\varphi|\pm E/v_F|\varphi|, \label{expansion2}
\end{align}
\end{subequations}
where $v_F=k_F/m$ is the Fermi velocity. Given the expansions in Eq.~(\ref{QCexp}), the condition for the solvability of Eq.~(\ref{BCs}) is
\\
\begin{equation} \label{QCsol}
\Omega\cos\left(\frac{2Ed}{v_F|\varphi|}\right)=E\sin\left(\frac{2Ed}{v_F|\varphi|}\right).
\end{equation}
It is then straightforward to solve explicitly for $\varphi$,
\begin{equation} \label{QCphi}
|\varphi_n|=\frac{2Ed/v_F}{\tan^{-1}(\Omega/E)+n\pi},
\end{equation}
where $n$ labels the de Gennes--Saint-James energy levels.

We consider the cases of thick ($d\gg d_c$) and thin ($d\ll d_c$) junctions separately, with $d_c$ as defined in Eq.~(\ref{dc}). In both cases, Eq.~(\ref{QCphi}) gives a solution $|\varphi_n|\sim Ed/v_F$ for $n>0$, while the $n=0$ level is
\begin{equation}
|\varphi_0|\sim\left\{\begin{array}{rcc}
	Ed/v_F, & & E\lesssim\Delta, \\
	(\Delta d/v_F)\sqrt{\Delta/(\Delta-E)}, & & E\approx\Delta, \end{array}\right.
\end{equation}
In order to satisfy condition (\ref{QCcond}) for the $|\varphi|\sim Ed/v_F$ solutions, we require that $E\gg\sqrt{\Delta/md^2}$. In the limit of a thick junction, where $\sqrt{\Delta/md^2}\ll\Delta$, the quasiclassical approximation breaks down at low energies $E\ll\Delta$. In the limit of a thin junction, where $\sqrt{\Delta/md^2}\gg\Delta$, we see that all solutions $|\varphi|\sim Ed/v_F$ are invalid for energies $E<\Delta$. The only valid solution in this limit is the $n=0$ solution for $E\approx\Delta$; condition (\ref{QCcond}) restricts the range of validity of this solution to a narrow interval near the bulk gap: $\Delta-E\ll\Delta^2md^2\ll\Delta.$ Thus, for both thin and thick junctions, the quasiclassical approximation breaks down below a certain energy. As will be shown in the rest of the paper, the spectrum is gapped below this energy scale.

\paragraph*{Quantum-mechanical treatment.}
The starting point for our fully quantum-mechanical treatment of the proximity effect is the exact solvability condition of Eq.~(\ref{BCs}), which can be expressed as
\begin{equation} \label{solvability}
f(k_\parallel)=0,
\end{equation}
with the dimensionless function $f(k_\parallel)$ given by \cite{supp}
\begin{widetext}
\begin{equation} \label{f}
\begin{aligned}
\Delta f(k_\parallel)&=\Omega\bar k_{+}\bar k_{-}\cos(k_+d)\cos(k_-d)+\Omega\left[w^2-iws(\bar p_+-\bar p_-)+s^2\bar p_+\bar p_-\right]\sin(k_+d)\sin(k_-d) \\
	&+\left[\Omega w\bar k_--E(\bar p_+u_0^2+\bar p_-v_0^2)\bar k_-\right]\sin(k_+d)\cos(k_-d)+\left[\Omega w\bar k_++E(\bar p_-u_0^2+\bar p_+v_0^2)\bar k_+\right]\cos(k_+d)\sin(k_-d).
\end{aligned}
\end{equation}
\end{widetext}
In Eq.~(\ref{f}), we introduce the dimensionless barrier strength $w=2U/v_{FN}$ and the Fermi velocity mismatch parameter $s=v_{FS}/v_{FN}$. We also define the dimensionless momenta $\bar p_\pm=p_\pm/k_{FS}$ and $\bar k_\pm=k_\pm/k_{FN}$. The proximity-induced gap $E_g$ is defined as the minimum energy for which a solution to Eq.~(\ref{solvability}) exists. While it is straightforward to determine $E_g$ numerically, we also examine several different limits analytically.

\begin{figure}[b!]
\includegraphics[width=0.75\linewidth]{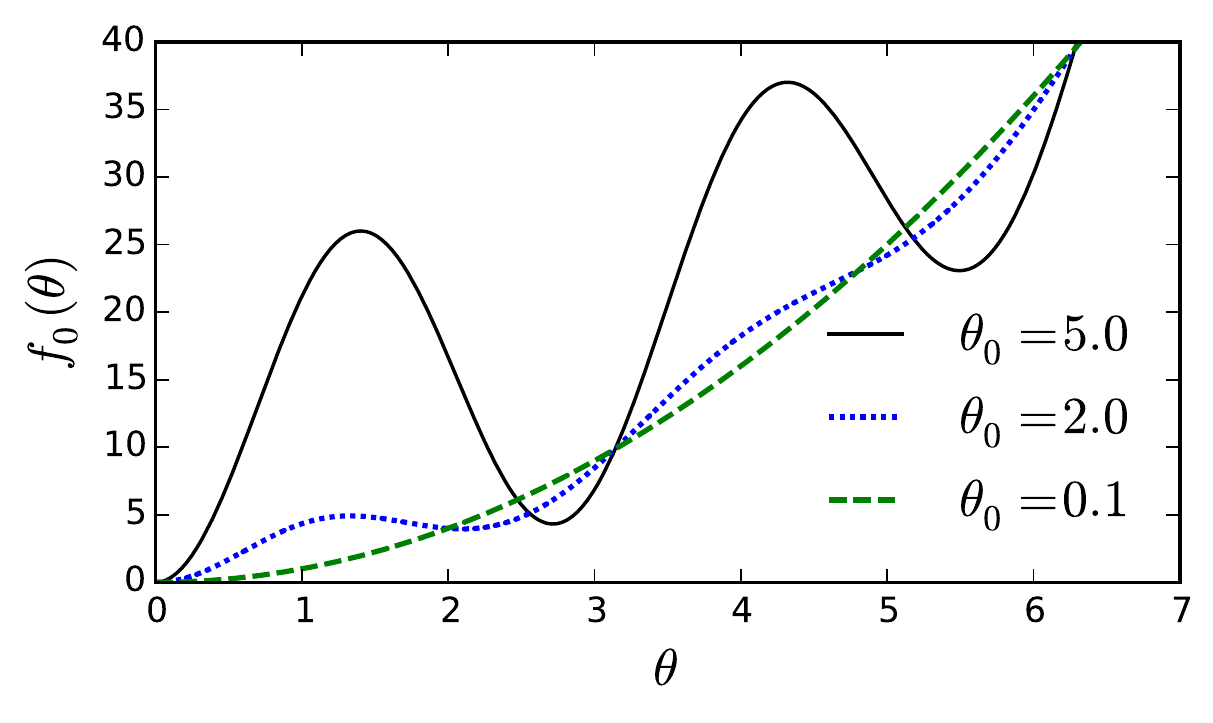}
\caption{\label{ftheta} (Color online) Plot of $f_0(\theta)$ at $E=0$ [Eq.~(\ref{f0})] for several values of $\theta_0=2\sqrt{\pi}(d/d_c)$. Because no solution to $f_0(\theta)=0$ exists, the normal layer is gapped for each $\theta_0$.}
\end{figure}

\paragraph*{No mismatch, no barrier.} We first revisit the case discussed previously in the context of the quasiclassical approximation, when there is neither Fermi surface mismatch ($E_{FN}=E_{FS},m_N=m_S$) nor an interfacial barrier ($w=0$). To show that a gap is induced for any value of $d$, we put $E=0$ directly in Eq.~(\ref{f}). With $f_0\equiv (k_Fd)^2f$, $\theta\equiv k_Fd\varphi$, and $\theta_0\equiv \sqrt{2\Delta md^2}=2\sqrt{\pi}(d/d_c)$,
\begin{equation} \label{f0}
\begin{aligned}
f_0(\theta)&=\theta^2\cos^2\theta+\sqrt{\theta^4+\theta_0^4}\sin^2\theta \\
&+\frac{\theta}{\sqrt{2}}\sqrt{\sqrt{\theta^4+\theta_0^4}-\theta^2}\sin2\theta.
\end{aligned}
\end{equation}
If no solution to $f_0(\theta)=0$ exists (aside from the trivial solution $\theta=0$, which corresponds to the wave function being identically zero in the normal layer), then $E=0$ is absent from the excitation spectrum and the system is gapped. Since $f_0(\theta)$ is an oscillatory function with $f_0(0)=0$ and $f_0'(0)>0$, a solution to $f_0(\theta)=0$ exists only if there is a local minimum of $f_0$ that is  negative. For a thin junction ($\theta_0\ll 1$),  $f_0(\theta)\approx \theta^2$ is a monotonically increasing function, and the spectrum is gapped. For a thick junction ($\theta_0\gg 1$), the function $f_0(\theta)\approx \theta^2\cos^2\theta+\theta_0^2\sin^2\theta+(\theta\theta_0/\sqrt{2})\sin2\theta$ has minima at $\theta^{\min}_n\approx n\pi\left(1-1/\sqrt{2}\theta_0\right)$ where $f_0(\theta^{\min}_n)\approx (n\pi)^2/2>0$, and thus the spectrum is gapped again. The function $f_0(\theta)$ for several values of $\theta_0$, including intermediate values $\theta_0\sim1$, is plotted in Fig.~\ref{ftheta}, showing that the spectrum is gapped for any choice of $\theta_0$. The magnitude of the gap ($E_g$) is determined as the minimum energy at which Eq.~(\ref{solvability}) has a solution.

It is natural to assume that $E_g\ll1/md^2$ for a thick junction. In this limit, the form of $k_{\pm}$  in Eq.~(\ref{expansion2}) still remains valid, while $p_{\pm}$ must be expanded in the limit opposite that of the quasiclassical approximation: $p_\pm^2/k_F^2\approx \pm i\Delta/E_F$. With these approximations, we obtain a minimum value $f_0(\theta^{\min}_1)=\pi^2/2-(\theta_0^6/4\pi^2)(E/\Delta)^2$, from which the gap is read off as
\begin{equation} \label{gap1}
E_g=\sqrt{2}\pi^2\frac{\Delta}{\theta_0^3}=\frac{\pi^2}{2md^2}\frac{1}{\sqrt{\Delta md^2}}.
\end{equation}
While Eq.~(\ref{gap1}) predicts that the gap is finite as long as $d$ finite, this result becomes irrelevant if the gap is very small. 
One obvious scale that $E_g$ needs to be compared with is the temperature; the other one is the minigap, $E_{\text{mg}}$, which
the quasiclassical theory predicts to open in a disordered normal layer. In the ballistic limit, $E_{\text{mg}}\sim1/\tau$ \cite{Reeg:2014,Pilgram:2000,Beenakker:2005}; in the diffusive limit, $E_{\text{mg}}\sim v_F^2\tau/d^2$ \cite{Pilgram:2000,Beenakker:2005,Golubov:1988,Belzig:1996,Altland:1998}, where $\tau$ is the scattering time. With even a small amount of disorder, the minigap is likely to be larger than the asymptotic limit given by Eq.~(\ref{gap1}).

For a thin junction, the forms of $p_\pm$ and $k_\pm$ given in Eq.~(\ref{QCexp}) remain valid. Expanding to leading order, we obtain $\Delta f_0(\theta)\approx\Omega\theta^2$. Since $f_0(\theta)$ is a monotonically increasing function in this limit, a solution to Eq.~(\ref{solvability}) exists only in the limit $\Omega\to0$. Therefore, the full bulk gap, $E_g\approx\Delta$, is induced in a thin junction.

The crossover between the two regimes occurs at $d\sim d_c$, with $d_c$ defined in Eq.~(\ref{dc}).

\begin{figure}[t!]
\includegraphics[width=.9\linewidth]{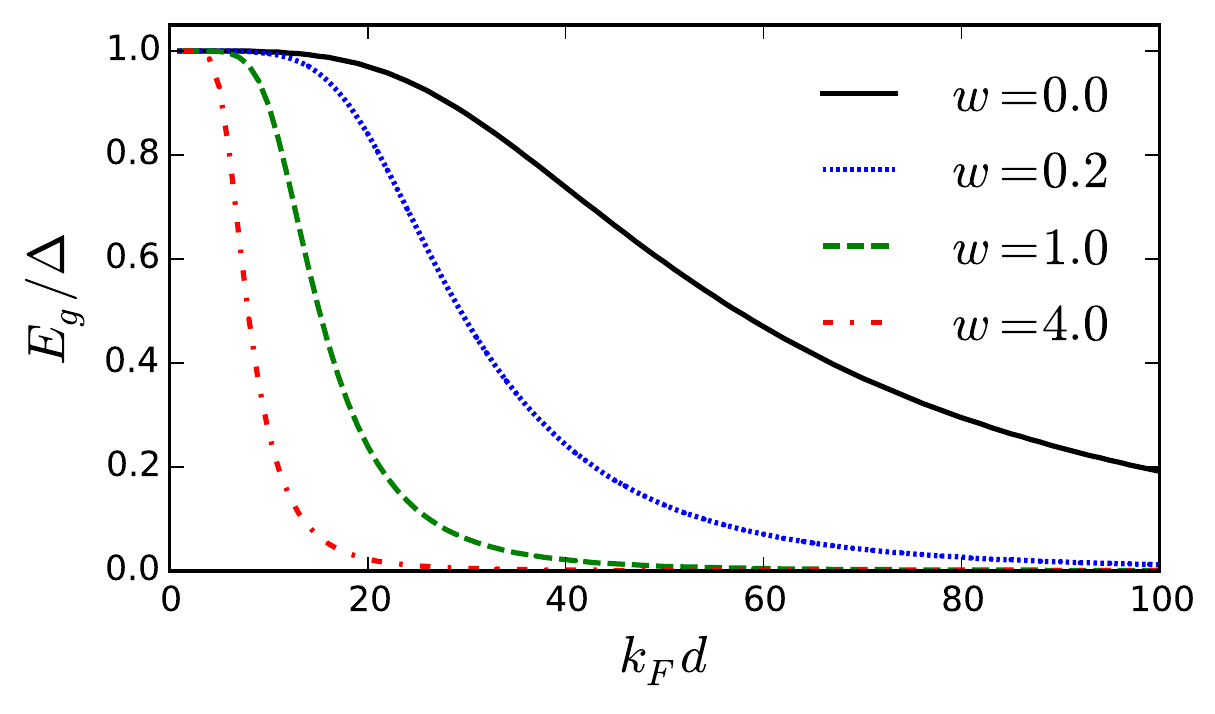}
\caption{\label{gap} (Color online) Numerical solution for the proximity-induced gap $E_g$ as a function of thickness in the absence of Fermi surface mismatch, plotted for various values of barrier strength $w$. Fermi energy was chosen so that $E_F/\Delta=10^3$.}
\end{figure}

\paragraph*{No mismatch, strong barrier.} We now consider the effect of an interfacial barrier on the induced gap. Anticipating that the barrier will decrease the gap, we focus only on the limit of a thin normal layer ($\Delta md^2\ll1$). The limit of a strong barrier can be treated analytically; the ``strong" barrier regime is defined by $w\gg1/k_Fd$, so that the $w^2$ term in Eq.~(\ref{f}) gives the leading contribution to $f(k_\parallel)$ for $\varphi\sim1/k_Fd$. In this regime, the gap is determined by the competition between  two large parameters: $wk_Fd$ and $1/md^2\Delta$ \cite{supp}. In the limit $w k_Fd\ll1/\Delta md^2$, the full bulk gap of the superconductor is again induced in the normal layer. In the opposite limit, only a small fraction of the bulk gap is induced,
\begin{equation}
E_g=\frac{\pi^3}{w^2(k_Fd)^2}\frac{1}{md^2}\ll\Delta.
\end{equation}
While it is still possible to induce a sizable gap in the presence of a strong barrier, the normal layer must be much thinner than $d_c$ as given in Eq.~(\ref{dc}):
\begin{equation} \label{crossover2}
d\lesssim d_c\left(\frac{\lambda_F}{\xi_S}\frac{1}{w^2}\right)^{1/4}\ll d_c,
\end{equation}
In Fig.~\ref{gap}, we plot a numerical solution of $E_g(d)$ for several values of $w$ in the limit of no Fermi surface mismatch.

\begin{figure}[t!]
\includegraphics[width=.9\linewidth]{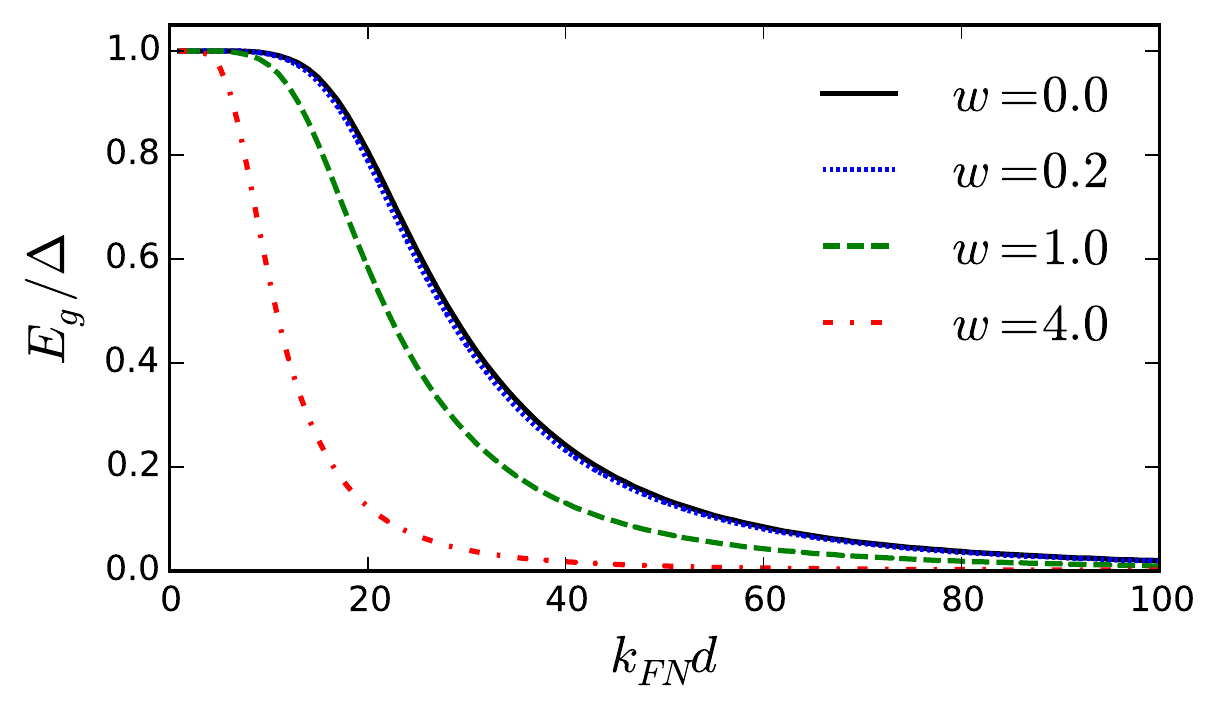}
\caption{\label{gap2} (Color online) Numerical solution for proximity-induced gap $E_g$ as a function of thickness with strong Fermi surface mismatch, plotted for various values of barrier strength $w$.  To enable a comparison with Fig.~\ref{gap}, we keep $E_{FN}/\Delta=10^3$ and $s=1$; we also choose $E_{FS}/E_{FN}=m_S/m_N=10$.}
\end{figure}

\paragraph*{Strong mismatch, no barrier.} 
We now consider the limit of strong Fermi surface mismatch. Having in mind a quasi-2D semiconductor quantum well coupled  to a superconductor, we consider the case when $k_{FN}\ll k_{FS}$ and $E_{FN}\ll E_{FS}$.

Focusing on thin junctions where $\Delta m_Nd^2\ll1$, we find that the induced gap is comparable to $\Delta$ provided that the Fermi velocity mismatch, which acts as an effective potential barrier at the interface, is sufficiently weak \cite{supp},
\begin{equation} \label{s}
\Delta m_Nd^2\ll s(k_{FN}d)\ll 1/\Delta m_Nd^2.
\end{equation}
Accounting for the fact that $s\gtrsim1$ in typical semiconductor/superconductor junctions, we find that
a large ($\sim \Delta$) proximity gap is induced provided that 
\begin{equation} \label{crossover3}
d\lesssim d_c \left(\lambda_{FN}/\xi_S s^4\right)^{1/6}\ll d_c,
\end{equation}
where $d_c$ is given by Eq.~(\ref{dc}) with  $\lambda_F$ replaced by $\lambda_{FN}$.

\paragraph*{Strong mismatch, strong barrier} 
Finally, we consider the case when both strong Fermi surface mismatch and a strong barrier ($w\gg1/k_{FN}d$) are present. Similarly to the case of no Fermi surface mismatch, the size of the induced gap is again determined by the competition between two large parameters \cite{supp}. When $1/\Delta m_Nd^2\gg(w^2+s^2)(k_{FN}d)/s$, the full bulk gap is induced in the normal layer; in the opposite limit, the induced gap is small,
\begin{equation} \label{gapcomp}
E_g=\frac{\pi^2}{m_Nd^2}\frac{s}{(w^2+s^2)(k_{FN}d)}\ll\Delta.
\end{equation}
We note that our result [Eq.~(\ref{gapcomp})] coincides with that of Ref.~\cite{Volkov:1995} in the 2D limit, when  $k_{FN}d\sim 1$. The $1/d^3$ scaling of our result is also in agreement with both Refs.~\cite{Fagas:2005} and \cite{Tkachov:2005}. A large ($\sim\Delta$) gap is induced if 
\begin{equation} \label{crossover4}
d\lesssim d_c \left[\frac{\lambda_{FN}}{\xi_S}\frac{1}{\left(w^2+s^2\right)^2}\right]^{1/6}\ll d_c.
\end{equation}
One interesting difference compared to the limit of no mismatch is that the combination of length scales $\bigl[d_c(\lambda_{FN}/\xi_S)^{1/6}\bigr]$ does not change in the presence of a barrier. As a result, the effect of a moderate barrier is actually weaker when there is strong mismatch; this can be seen clearly in Fig.~\ref{gap2}, which plots a numerical solution of $E_g(d)$ for various values of $w$ in the strong mismatch limit.

\paragraph*{Conclusion.} We have shown that a hard superconducting gap is proximity-induced in a normal layer of any finite thickness and have studied the dependence of this gap on the thickness of the normal layer. It is possible to induce a sizable fraction of the full bulk gap of the superconductor in layers that are much thicker than the Fermi wavelength, a result that is relatively robust to moderate interfacial barrier strengths and strong Fermi surface mismatch. The analytic results for the crossover thickness, below which the induced gap is comparable to the bulk gap of the superconductor, are summarized in Table~\ref{table}. 

The ability to induce a superconducting gap via the proximity effect has been well demonstrated experimentally. Gaps observed in tunneling experiments on mesoscopic junctions \cite{Gueron:1996,Moussy:2001,SerrierGarcia:2013} can be attributed to the diffusive nature of the normal layer and correspond to the disorder-induced minigap. Conversely, in nanoscale junctions involving either InAs or InSb nanowires, gaps observed in transport experiments probing topological superconductivity \cite{Mourik:2012,Das:2012,Deng:2012,Finck:2013} can be attributed to the finite-size effects discussed in this paper. In many materials the observed gap appears ``soft"; i.e., there remains a finite density of states at the Fermi energy. However, there have been recent observations of a hard superconducting gap \cite{Chang:2015,Kjaergaard:2016,Zhang:2016}, an important step toward developing Majorana-based quantum devices. As a sizable gap is needed to stabilize topological superconductivity, the results of this paper significantly lessen the experimental restrictions on the thickness of the proximity-coupled layer in order to induce such a gap.

\begin{table}[t!]
\begin{tabular*}{\linewidth}{ @{\extracolsep{\fill}} l  c  c }
	\hline \hline
	  & No barrier & Strong barrier \\
	 \hline
	No mismatch & $\displaystyle 1$ & $\displaystyle \left(\frac{\lambda_F}{\xi_Sw^2}\right)^{1/4}$ \\[0.5cm]
	
	Strong mismatch & $\displaystyle \left(\frac{\lambda_{FN}}{\xi_Ss^4}\right)^{1/6}$ & $\displaystyle \left[\frac{\lambda_{FN}}{\xi_S(w^2+s^2)^2}\right]^{1/6}$ \\[0.3cm]
	\hline \hline
\end{tabular*}
\caption{\label{table} 
Crossover thickness of the normal layer in units of $d_c$ defined by Eq.~(\ref{dc}). If $d$ is less than the corresponding thickness, an induced gap is comparable to the bulk gap in the superconductor.}
\end{table}

\paragraph*{Acknowledgments.}
We thank C. Batista, O. Entin-Wohlman, A. Kumar, S. Lin, S. Maiti, I. Martin, M. Metlitski, and V. Yudson for useful discussions. This work was supported by the National Science Foundation via grant NSF DMR-1308972. We also acknowledge the hospitality of the Center for Nonlinear Studies, Los Alamos National Laboratory, where a part of this work was done.

\bibliography{bibProximityGap}

\begin{widetext}

\setcounter{equation}{0}
\setcounter{figure}{0}
\renewcommand{\theequation}{S\arabic{equation}}
\renewcommand{\thefigure}{S\arabic{figure}}

\section{Supplementary Material for \\ ``Hard gap in a normal layer coupled to a superconductor"}

\subsection{Solution strategy}
The boundary conditions given in Eq.~(\ref{BCs}) of the main text form a system of four equations that must be solved simultaneously. In matrix form, this system of equations is given by
\begin{equation} \label{BCmatrix}
\left(\begin{array}{cccc}
	u_0 & v_0 & -\sin k_+d & 0 \\
	v_0 & u_0 & 0 & -\sin k_-d \\
	iu_0u_{+} & -i v_0u_{-} & -v_{+}\cos(k_+d)-2U\sin(k_+d) & 0 \\
	iv_0 u_{+} & -iu_0u_{-} & 0 & -v_{-}\cos(k_-d)-2U\sin(k_-d)
	\end{array}\right)\left(\begin{array}{c} c_1 \\ c_2 \\ c_3 \\ c_4 \end{array}\right)=0,
\end{equation}
where we define the velocities $u_\pm=p_\pm/m_S$ and $v_\pm=k_\pm/m_N$. The condition for the solvability of Eq.~(\ref{BCmatrix}) is
\begin{equation} \label{solvability}
f(\varphi_N)=0,
\end{equation}
where we reexpress the definition of $f(\varphi_N)$ given in Eq.~(\ref{f}) of the main text as
\begin{equation} \label{fsupp}
\begin{aligned}
\Delta f(\varphi_N)&=\Omega\sqrt{\varphi_N^4-E^2/E_{FN}^2}\cos(k_+d)\cos(k_-d)+\Omega\left[w^2+\frac{\sqrt{2}ws(\Omega/E_{FS})}{\sqrt{\varphi_S^2+\sqrt{\varphi_S^4+\Omega^2/E_{FS}^2}}}+s^2\sqrt{\varphi_S^4+\Omega^2/E_{FS}^2}\right]\sin(k_+d)\sin(k_-d) \\
	&+\left[\Omega w\sqrt{\varphi_N^2+E/E_{FN}}+\frac{s}{\sqrt{2}}\left(E\varphi_S^2+\frac{\Omega^2}{E_{FS}}+E\sqrt{\varphi_S^4+\Omega^2/E_{FS}^2}\right)\sqrt{\frac{\varphi_N^2+E/E_{FN}}{\varphi_S^2+\sqrt{\varphi_S^4+\Omega^2/E_{FS}^2}}}\right]\cos(k_+d)\sin(k_-d) \\
	&+\left[\Omega w\sqrt{\varphi_N^2-E/E_{FN}}-\frac{s}{\sqrt{2}}\left(E\varphi_S^2-\frac{\Omega^2}{E_{FS}}+E\sqrt{\varphi_S^4+\Omega^2/E_{FS}^2}\right)\sqrt{\frac{\varphi_N^2-E/E_{FN}}{\varphi_S^2+\sqrt{\varphi_S^4+\Omega^2/E_{FS}^2}}}\right]\sin(k_+d)\cos(k_-d).
\end{aligned}
\end{equation}
Note that $f(\varphi_N)$ is a function of only a single variable parameterized by the in-plane momentum $k_\parallel$, as we can relate $\varphi_S^2=1-(k_{FN}/k_{FS})^2(1-\varphi_N^2)$. If at a given energy $E$ there does not exist a value of $\varphi_N$ that solves Eq.~(\ref{solvability}), then this energy is absent from the excitation spectrum of the normal layer and lies within the proximity-induced gap. The magnitude of the gap $E_g$ is defined to be the minimum energy for which a solution to Eq.~(\ref{solvability}) exists.

While the form given in Eq.~(\ref{fsupp}) is indeed very complicated, our plan of attack for determining the size of the gap analytically is informed by the general behavior of $f(\varphi_N)$, which is displayed in Fig.~\ref{fplot}. Two important properties of $f(\varphi_N)$ are immediately apparent upon examining these plots. First, $\varphi_N^2=E/E_{FN}$ always solves Eq.~(\ref{solvability}); however, this choice corresponds to $k_-=0$ and represents a trivial solution. We therefore search for solutions that satisfy $\varphi_N^2>E/E_{FN}$. Second, it is clear that $E_g$ can be identified as the energy at which the first minimum in $f(\varphi_N)$ goes to zero. Accordingly, our analytical strategy for determining the size of the gap will be to first determine the value $\varphi_\text{min}$ corresponding to this minimum before solving for the energy at which $f(\varphi_\text{min})=0$.

\subsection{No mismatch, no barrier}
The first case we will consider is that of no Fermi surface mismatch ($m_N=m_S,E_{FN}=E_{FS},\varphi_N=\varphi_S,s=1$) and no interfacial barrier ($w=0$). Given that the arguments of the oscillatory factors present in Eq.~(\ref{fsupp}) are of the form
\begin{equation} \label{arg}
k_\pm d=k_Fd\sqrt{\varphi^2\pm E/E_F},
\end{equation}
the function $f(\varphi)$ oscillates on a scale $\varphi\sim 1/k_Fd$ when $\varphi^2\gg E/E_{F}$. Therefore, probing the first minimum $\varphi_\text{min}\sim1/k_Fd$ allows an expansion of Eq.~(\ref{arg}) in two different limits: $\Delta md^2\ll1$ (equivalently, $k_F^2d^2\ll E_F/\Delta$) and $\Delta md^2\gg1$ (equivalently, $k_F^2d^2\gg E_F/\Delta$). We will now examine these two limits separately.

In the limit $\Delta md^2\ll1$, we can expand Eq.~(\ref{fsupp}) for $\varphi^2\gg\Delta/E_F$. Expanding Eq.~(\ref{fsupp}) to leading order, we find
\begin{equation}
\Delta f(\varphi\sim1/k_Fd)=\Omega\varphi^2.
\end{equation}
Because this is a monotonically increasing function, the only way to satisfy Eq.~(\ref{solvability}) is to have $\Omega=0$, or
\begin{equation} \label{fullgap}
E_g=\Delta.
\end{equation}
Therefore, the full bulk gap of the superconductor is induced in the normal layer.

\begin{figure}[t!]
\includegraphics[width=.5\linewidth]{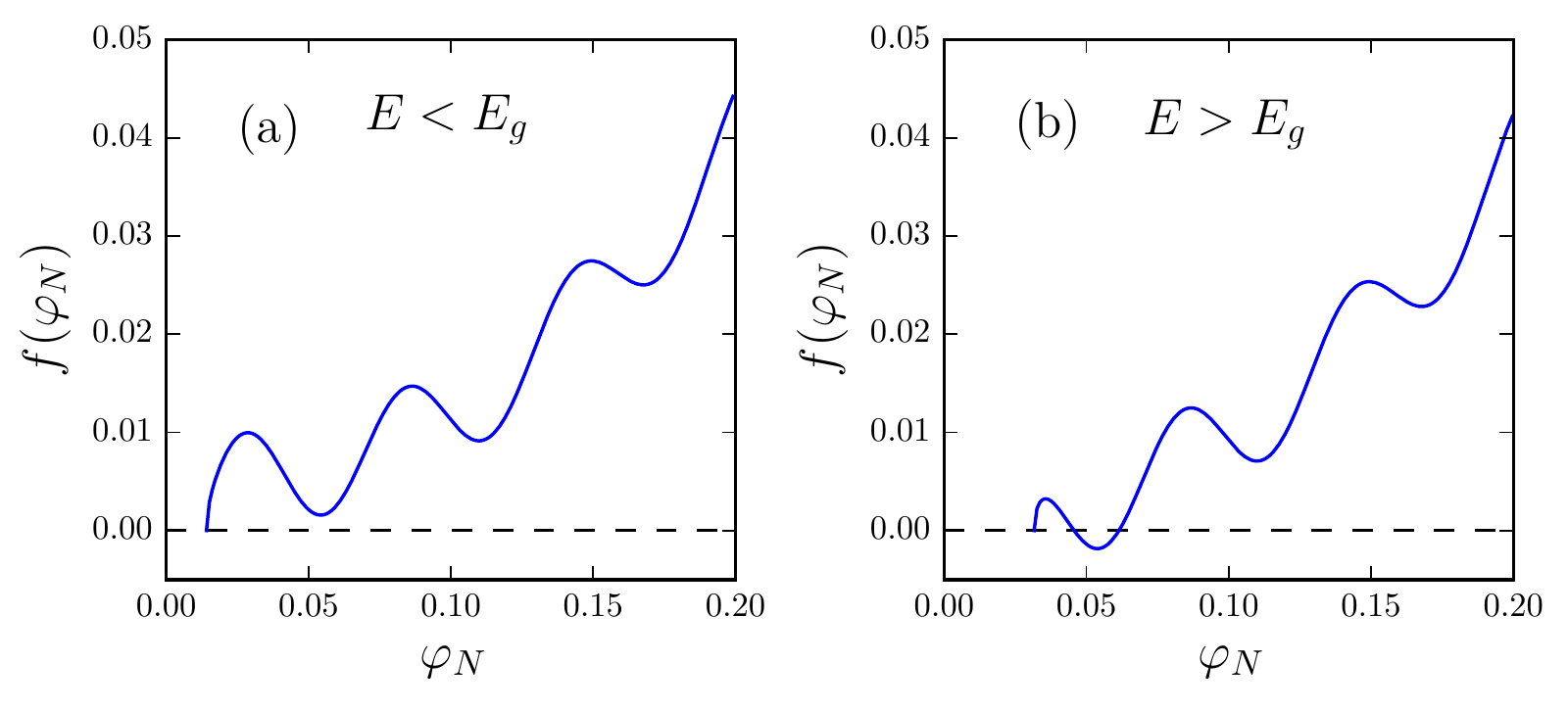}
\caption{\label{fplot} (Color online) (a) For energies below the gap ($E<E_g$), no solution to $f(\varphi_N)=0$ exists. (b) At least one solution to $f(\varphi_N)=0$ exists for energies above the gap ($E>E_g$).}
\end{figure}

We now consider the opposite limit, where $\Delta md^2\gg1$. Anticipating that a small gap will be induced in the normal layer, we consider energies that satisfy $E\ll1/md^2$. This assumption allows us to expand for $E/E_F\ll\varphi^2\ll\Delta/E_F$. Expanding Eq.~(\ref{fsupp}), and replacing $\Omega\to\Delta$, gives
\begin{equation} \label{expansion1}
f(\varphi\sim1/k_Fd)=\frac{\Delta}{E_F}\left(\sin^2(k_Fd\varphi)-\frac{(k_Fd)^2E^2}{4E_F^2\varphi^2}\right)+\frac{\varphi}{\sqrt{2}}\sqrt{\frac{\Delta}{E_F}}\sin(2k_Fd\varphi)+\varphi^2\cos^2(k_Fd\varphi).
\end{equation}
While the first term in Eq.~(\ref{expansion1}) represents the leading term in the expansion, all given terms are comparable in magnitude in the vicinity of $\varphi_\text{min}$. The leading contribution to $\varphi_\text{min}$ is determined by the vanishing of the first term in Eq.~(\ref{expansion1}), $\varphi_\text{min}\approx\pi/k_Fd$. To calculate the first-order correction to this value, we take $\varphi_\text{min}=\pi(1-\delta\varphi)/k_Fd$, with $\delta\varphi\ll1$, and expand in the vicinity of $\varphi_\text{min}$ to obtain (any subleading terms are neglected here)
\begin{equation}
f'(\varphi_\text{min})=-2\pi (k_Fd)\delta\varphi (\Delta/E_F)+\pi\sqrt{2\Delta/E_F}.
\end{equation}
Solving for the minimum, where $f'(\varphi_\text{min})=0$, we find
\begin{equation}
\varphi_\text{min}=\frac{\pi}{k_Fd}\left(1-\frac{\sqrt{E_F/2\Delta}}{k_Fd}\right).
\end{equation}
Expanding Eq.~(\ref{expansion1}) in the vicinity of $\varphi_\text{min}$, we obtain
\begin{equation}
f(\varphi_\text{min})=\frac{1}{2}\left(\frac{\pi}{k_Fd}\right)^2-\frac{(k_Fd)^4\Delta E^2}{4\pi^2E_F^3}.
\end{equation}
The gap is defined to be the energy at which $f(\varphi_\text{min})=0$. Solving for $E$, we find an expression for the gap given by
\begin{equation}
E_g=\frac{\pi^2}{2md^2}\frac{1}{\sqrt{\Delta md^2}}.
\end{equation}
In contrast to the previous case, only a very small fraction of the full superconducting gap is induced in a sufficiently thick normal layer.


\subsection{No mismatch, strong barrier}
In this section, we calculate the gap in the presence of a strong barrier. Focusing on the limit $\Delta md^2\ll1$, we expand Eq.~(\ref{fsupp}) for $\varphi^2\gg\Delta/E_F$ to give
\begin{equation} \label{barrierf}
\begin{aligned}
\Delta f(\varphi)&=\Omega\left[\varphi^2+\left(w^2+\frac{\Omega w}{E_F\varphi}\right)\left(\sin^2(k_Fd\varphi)-\frac{(k_Fd)^2E^2}{4E_F^2\varphi^2}\right)+\left(w\varphi+\frac{\Omega}{2E_F}\right)\sin(2k_Fd\varphi)\right]-\frac{E^2}{E_F}k_Fd\varphi.
\end{aligned}
\end{equation}
As we did previously, we keep all terms that are relevant for determining either $\varphi_\text{min}$ or $f(\varphi_\text{min})$. In order for the term proportional to $w^2$ to form the leading contribution to $f(\varphi)$, the barrier must be strong enough to satisfy $w\gg 1/k_Fd$. In this case, the leading contribution to $\varphi_\text{min}$ is again given by $\varphi_\text{min}\approx\pi/k_Fd$. To find the first-order correction, we write $\varphi_\text{min}=\pi(1-\delta\varphi_1)/k_Fd$ and expand for $\delta\varphi_1\ll1$, giving
\begin{equation}
\Delta f'(\varphi_\text{min})=\Omega\bigl[(-2\pi w^2k_Fd)\delta\varphi_1+2\pi w\bigr].
\end{equation}
Solving for the minimum gives $\delta\varphi_1=1/w(k_Fd)$. However, when expanding Eq.~(\ref{barrierf}) in the vicinity of $\varphi_\text{min}$ to order $\mathcal{O}(1/k_F^2d^2)$, we find that all terms independent of $E$ cancel. We therefore must go beyond first order in determining $\varphi_\text{min}$. Writing $\varphi_\text{min}=\pi(1-1/wk_Fd+\delta\varphi_2)/k_Fd$ and expanding for $\delta\varphi_2\ll1/wk_Fd$, we obtain
\begin{equation}
\Delta f'(\varphi_\text{min})=\Omega\left[-\frac{2\pi}{k_Fd}+(2\pi w^2k_Fd)\delta\varphi_2\right].
\end{equation}
Solving for the minimum gives a second-order correction of $\delta\varphi_2=1/(wk_Fd)^2$. Now expanding Eq.~(\ref{barrierf}) to order $\mathcal{O}(1/wk_F^3d^3)$, we again find a cancelation of all terms independent of $E$. Going still further in the expansion for $\varphi_\text{min}$, we write $\varphi_\text{min}=\pi(1-1/wk_Fd+1/w^2k_F^2d^2+\delta\varphi_3)/k_Fd$ and expand for $\delta\varphi_3\ll 1/(wk_Fd)^2$,
\begin{equation}
\Delta f'(\varphi_\text{min})=(2\pi w^2k_Fd)\delta\varphi_3-\frac{8\pi^3}{3w(k_Fd)^2}+\frac{2\pi}{w(k_Fd)^2}-\frac{\Omega}{E_F}k_Fd.
\end{equation}
Solving for the minimum, we find $\delta\varphi_3=(4\pi^2/3-1)/(wk_Fd)^3+\Omega/2\pi E_Fw^2$. Combining all orders, we have
\begin{equation}
\varphi_\text{min}=\frac{\pi}{k_Fd}\left(1-\frac{1}{wk_Fd}+\frac{1}{(wk_Fd)^2}+\frac{4\pi^2/3-1}{(wk_Fd)^3}+\frac{\Omega}{2\pi E_Fw^2}\right).
\end{equation}
Expanding Eq.~(\ref{barrierf}) to order $\mathcal{O}(1/w^2k_F^4d^4)$ gives
\begin{equation} \label{barrierf2}
\Delta f(\varphi_\text{min})=\Omega\left[\frac{\pi^4}{w^2(k_Fd)^4}-\frac{w^2(k_Fd)^4E^2}{4\pi^2E_F^2}\right]-\frac{\pi E^2}{E_F}.
\end{equation}

The dominant $E^2$ term in Eq.~(\ref{barrierf2}) is determined by the strength of the barrier. In the limit $w^2(k_Fd)^2\gg1/\Delta md^2\gg1$, the second $E^2$ term can be neglected compared to the first. It is then very straightforward to solve for the gap:
\begin{equation}
E_g=\frac{\pi^3}{w^2(k_Fd)^2}\frac{1}{md^2}\ll\Delta.
\end{equation}
In the opposite limit, where $1/\Delta md^2\gg w^2(k_Fd)^2\gg1$, the first $E^2$ term can be neglected and we have
\begin{equation} \label{barrierf3}
\Delta f(\varphi_\text{min})=\frac{\pi^4\Omega}{w^2(k_Fd)^4}-\frac{\pi E^2}{E_F}.
\end{equation}
If $\Omega\sim\Delta$, then the first term in Eq.~(\ref{barrierf3}) is always much larger in magnitude than the second. The two terms can only be comparable if $\Omega\ll\Delta$; this indicates that $E_g\approx\Delta$. To calculate the small correction to the gap, we can express $E_g=\Delta-\delta E$ and replace $\Omega=(2\Delta\delta E)^{1/2}$ in Eq.~(\ref{barrierf3}). Solving $f(\varphi_\text{min})=0$ for $\delta E$, we find that the proximity-induced gap is given by
\begin{equation}
E_g=\Delta\left(1-\frac{2}{\pi^6}\left[w^2(k_Fd)^2\Delta md^2\right]^2\right)\approx\Delta.
\end{equation}

\subsection{Strong mismatch, no barrier}
In this section, we consider the limit of strong Fermi surface mismatch, so that $k_{FN}\ll k_{FS}$ and $E_{FN}\ll E_{FS}$. Because the in-plane momentum $k_\parallel$ has an upper limit of $k_{FN}$, we can approximate $\varphi_S=1$. In the limit $\Delta m_Nd^2\ll1$, we can expand Eq.~(\ref{fsupp}) for $\varphi_S^2\geq\varphi_N^2\gg\Delta/E_{FN}\gg \Delta/E_{FS}$ because the oscillation scale of $f(\varphi_N)$ is set by $\varphi_N\sim1/k_{FN}d$. To leading order, we have
\begin{equation} \label{fmismatch}
\Delta f(\varphi_N\sim1/k_{FN}d)=\Omega\left[\varphi_N^2\cos^2(k_{FN}d\varphi_N)+s^2\sin^2(k_{FN}d\varphi_N)+\frac{s\Omega\varphi_N}{2E_{FS}}\sin(2k_{FN}d\varphi_N)\right]
\end{equation}
As long as $\Delta/E_{FS}\ll s(k_{FN}d)\ll E_{FS}/\Delta$, the third term in Eq.~(\ref{fmismatch}) can be neglected compared to the first two terms. When $s(k_{FN}d)\sim1$, the first two terms in Eq.~(\ref{fmismatch}) are comparable in magnitude and $f(\varphi_N)$ is a positive-definite function; i.e., $f(\varphi_N)$ cannot be driven to zero at any $\varphi_N$ by corrections to Eq.~(\ref{fmismatch}). Therefore, the only solution satisfying $f(\varphi_N)=0$ is $E_g=\Delta$.

This argument breaks down, however, if the Fermi velocity mismatch is sufficiently strong. Let us consider the case where $\Delta/E_{FS}\ll s(k_{FN}d)\ll 1$; in this limit, we can expand Eq.~(\ref{fsupp}) beyond what is given in Eq.~(\ref{fmismatch}) to include relevant corrections,
\begin{equation} \label{fmismatch2}
\Delta f(\varphi_N\sim1/k_{FN}d)=\Omega\left[\varphi_N^2\cos^2(k_{FN}d\varphi_N)+s^2\sin^2(k_{FN}d\varphi_N)-\frac{E^2(k_{FN}d)^2}{4E_{FN}^2}\right]-\frac{E^2}{E_{FN}}s(k_{FN}d).
\end{equation}
In the vicinity of the first minimum of $f(\varphi_N)$, which is located near $\varphi_N=\pi/2k_{FN}d$, the second term in Eq.~(\ref{fmismatch2}) is much larger in magnitude than the two $E^2$ correction terms provided that $s(k_{FN}d)\gg \Delta m_Nd^2$. If this condition holds, then the only solution to $f(\varphi_N)=0$ must again be $E_g=\Delta$.

If we instead consider the case where $1\ll s(k_{FN}d)\ll E_{FS}/\Delta$, expanding Eq.~(\ref{fsupp}) beyond what is given in Eq.~(\ref{fmismatch}) gives
\begin{equation} \label{fmismatch3}
\Delta f(\varphi_N\sim1/k_{FN}d)=\Omega\left[\varphi_N^2\cos^2(k_{FN}d\varphi_N)+s^2\sin^2(k_{FN}d\varphi_N)-\frac{E^2s^2(k_{FN}d)^2}{4E_{FN}^2\varphi_N^2}\right]-\frac{E^2}{E_{FN}}s(k_{FN}d).
\end{equation}
In the vicinity of the first minimum of $f(\varphi_N)$, which is now located near $\varphi_N=\pi/k_{FN}d$, the first term in Eq.~(\ref{fmismatch3}) is much larger in magnitude than the two $E^2$ corrections terms provided that $s(k_{FN}d)\ll 1/m_Nd^2 \Delta$.
Once again, if this condition holds, the only solution to $f(\varphi_N)=0$ is $E_g=\Delta$.

\subsection{Strong Fermi surface mismatch, strong barrier}
Finally, we consider the case of strong Fermi surface mismatch and strong barrier, so that $w\gg 1/k_{FN}d$. Assuming that $\Delta m_Nd^2\ll1$, we can again expand Eq.~(\ref{fsupp}) to leading order for $\varphi_S^2\geq\varphi_N^2\gg\Delta/E_{FN}\gg\Delta/E_{FS}$. Keeping terms for $w\neq0$, we now find that
\begin{equation} \label{fmismatch4}
\Delta f(\varphi_N\sim1/k_{FN}d)=\Omega\left[\varphi_N^2\cos^2(k_{FN}d\varphi_N)+\left(w^2+s^2+\frac{ws\Omega}{E_{FS}}\right)\sin^2(k_{FN}d\varphi_N)+\varphi_N\left(w+\frac{s\Omega}{2E_{FS}}\right)\sin(2k_{FN}d\varphi_N)\right].
\end{equation}
Provided that $\Delta/E_{FS}\ll w/s\ll E_{FS}/\Delta$, both terms proportional to $\Omega/E_{FS}$ in Eq.~(\ref{fmismatch4}) can be neglected. Recognizing that the first minimum in $f(\varphi_N)$ can be expressed as $\varphi_\text{min}=\pi(1-\delta\varphi)/k_{FN}d$, for $\delta\varphi\ll1$, we can expand near $\varphi_\text{min}$ to obtain
\begin{equation}
\Delta f'(\varphi_\text{min})=\Omega\biggl\{\bigl[-2\pi(w^2+s^2)k_{FN}d\bigr]\delta\varphi+2\pi w\biggr\}.
\end{equation}
Solving for the minimum $f'(\varphi_\text{min})=0$, we find that
\begin{equation}
\varphi_\text{min}=\frac{\pi}{k_{FN}d}\left(1-\frac{w}{(w^2+s^2)k_{FN}d}\right).
\end{equation}
While so far we have kept $s\sim w$, we are justified in stopping at this first term in the expansion for $\varphi_\text{min}$ in the limit $s\ll w$ only if $s^2(k_{FN}d)^2\gg1$ (though details on how to show this are omitted); we will proceed under this assumption. Going back and expanding Eq.~(\ref{fsupp}) beyond leading order to include possible leading $E^2$ corrections,
\begin{equation} \label{fmismatch5}
\begin{aligned}
\Delta f(\varphi_N\sim1/k_{FN}d)&=\Omega\biggl[\varphi_N^2\cos^2(k_{FN}d\varphi_N)+(w^2+s^2)\left(\sin^2(k_{FN}d\varphi_N)-\frac{(k_{FN}d)^2E^2}{4E_{FN}^2\varphi_N^2}\right) \\
	&+w\varphi_N\sin(2k_{FN}d\varphi_N)\biggr]-\frac{E^2}{E_{FN}}s(k_{FN}d).
\end{aligned}
\end{equation}
In the vicinity of $\varphi_\text{min}$, we expand Eq.~(\ref{fmismatch5}) to give
\begin{equation} \label{fmismatch6}
\Delta f(\varphi_\text{min})=\Omega\left[\frac{\pi^2s^2}{(k_{FN}d)^2(w^2+s^2)}-(w^2+s^2)\frac{(k_{FN}d)^4E^2}{4\pi^2E_{FN}^2}\right]-\frac{E^2}{E_{FN}}s(k_{FN}d).
\end{equation}

In the limit $(w^2+s^2)(k_{FN}d)/s\gg1/\Delta m_Nd^2\gg1$, the first $E^2$ correction term is always much larger in magnitude than the second. Neglecting the second $E^2$ term, we solve for the gap to give
\begin{equation}
E_g=\frac{\pi^2}{m_Nd^2}\frac{s}{(w^2+s^2)(k_{FN}d)}\ll\Delta.
\end{equation}
In the opposite limit, where $1/\Delta m_Nd^2\gg(w^2+s^2)(k_{FN}d)/s\gg1$, the first $E^2$ correction term can be neglected. In this limit, the first term in Eq.~(\ref{fmismatch6}) is always much larger in magnitude than the third term unless $\Omega\ll\Delta$. This suggests that the full bulk gap of the superconductor is induced in the normal layer. Writing $E_g=\Delta-\delta E$, we have
\begin{equation}
\Delta f(\varphi_\text{min})=\frac{\pi^2s^2\sqrt{2\Delta\delta E}}{(k_{FN}d)^2(w^2+s^2)}-\frac{\Delta^2}{E_{FN}}s(k_{FN}d).
\end{equation}
Solving $f(\varphi_\text{min})=0$ for the correction $\delta E$, we find that the induced gap is given by
\begin{equation}
E_g=\Delta\left(1-\frac{2}{\pi^4}\left[\frac{1}{s}(w^2+s^2)(k_{FN}d)\Delta m_Nd^2\right]^2\right).
\end{equation}

\end{widetext}

\end{document}